\documentclass[aps,showpacs,amsmath,amssymb,prb,groupedaddress]{revtex4-1}
\usepackage{color}
\usepackage{graphicx}
\usepackage{dcolumn}
\usepackage{bm}
\newcommand{\eight}{{\ensuremath{\frac{1}{8}}}}

\begin{document}

\title{Nonconventional magnetic order in frustrated diamond lattice antiferromagnet CoAl$_2$O$_4$ studied by neutron diffraction and classical Monte-Carlo simulation}

\author{O. Zaharko}
\affiliation{Laboratory for Neutron Scattering and Imaging, Paul Scherrer Institute, CH-5232 Villigen, Switzerland}
\author{S. T\'oth}
\affiliation{Laboratory for Neutron Scattering and Imaging, Paul Scherrer Institute, CH-5232 Villigen, Switzerland}
\author{O. Sendetskyi}
\affiliation{Laboratory for Micro- und Nanotechnology, Paul Scherrer Institute, CH-5232 Villigen, Switzerland}
\author{A. Cervellino}
\affiliation{Laboratory for Synchrotron Radiation - Condensed Matter, Swiss Light Source, Paul Scherrer Institute, CH-5232 Villigen, Switzerland}
\author{A. Wolter-Giraud}
\affiliation{Leibniz Institute for Solid State and Materials Research Dresden, Helmholtzstrasse 20, 01069 Dresden, Germany}
\author{T. Dey}
\affiliation{Leibniz Institute for Solid State and Materials Research Dresden, Helmholtzstrasse 20, 01069 Dresden, Germany}
\author{A. Maljuk}
\affiliation{Leibniz Institute for Solid State and Materials Research Dresden, Helmholtzstrasse 20, 01069 Dresden, Germany}
\author{V. Tsurkan}
\affiliation{Experimental Physics V, Center for Electronics Correlations and Magnetism, University of Augsburg, D-86159 Augsburg, Germany}
\affiliation{Institute of Applied Physics, Academy of Sciences of Moldova, MD-2028 Chisinau, Republic of Moldova}

\date{\today}

\begin{abstract}
CoAl$_2$O$_4$ spinel with magnetic Co$^{2+}$ ions on the diamond A-lattice is known to be magnetically frustrated. 
We compare neutron single crystal diffraction patterns measured in zero and applied magnetic fields with the ones obtained from classical Monte-Carlo models. In simulations we test the influence of various parameters on diffraction patterns: the ratio of nearest-, $J_1$, and next-nearest, $J_2$, neighbor interactions, magnetic field applied along the principal crystallographic directions, and random disorder on the A(Co$^{2+}$)- and B(Al$^{3+}$)- sites. We 
conclude that the models considered so far explain the broadening of magnetic Bragg peaks in zero magnetic field and their anisotropic response to applied magnetic field only partly. As bulk properties of our single crystal are isotropic, we suggest that its microstructure, specifically $<$111$>$-twin boundaries, could be a reason of the nonconventional magnetic order in CoAl$_2$O$_4$.
\end{abstract}

\maketitle

\section{Introduction{\label{Sec1}}}
Theoretical studies\cite{Bergman07} predict that the magnetic geometrically nonfrustrated diamond lattice can host a highly degenerate frustrated state consisting of coplanar spirals, the so-called spiral spin liquid (SSL). 
This counterintuitive finding is comprehended when taking into account that the diamond lattice is composed of two interpenetrated face-centered cubic (FCC) sublattices which on their own are highly frustrated\cite{Gvozdikova05}. 
The magnetic frustration of the overall diamond lattice is caused by a significant J$_2$ interaction which
couples nearest-neighbor sites within each FCC sublattice.
The SSL should evolve\cite{Bergman07} when the ratio $J_2$/$J_1$ exceeds {\eight}, while for lower $J_2$/$J_1$ a collinear antiferromagnet (AF) should be the ground state. 
The labeling of J's reflects the fact that J$_1$ is the nearest-neighbor coupling of the diamond lattice and J$_2$ is the next-nearest-neighbor one.
The degeneracy of the SSL could be lifted by thermal\cite{Bergman07} or quantum\cite{Bernier08} fluctuations resulting in a magnetic ordering transition via the 'order-by-disorder' mechanism\cite{Villian80, Henley89}. Within this scenario the ordered state selected by entropy has the highest density of nearby low-energy excitations.
This model will be referred to as the '$J_2$/$J_1$' or 'order-by-disorder' model.\\ 
CoAl$_2$O$_4$ is a good candidate to study SSL, as it adopts the spinel structure AB$_2$O$_4$ with magnetic Co$^{2+}$ ions residing on the diamond lattice. However, there is no consensus to which extent 
the $J_2$/$J_1$ model is relevant for this system. 
The ratio $J_2$/$J_1$ determined from inelastic neutron scattering\cite{Zaharko11} is only 0.109(2), thus, according to the aforementioned model\cite{Bergman07}, this material should have a nondegenerate collinear AF ground state. Nevertheless, experimental observations by specific heat\cite{Tristan05,Tristan08}, muon spin relaxation\cite{Suzuki07}, neutron diffraction\cite{Zaharko10, Zaharko11, MacDougall11} are rather consistent with an unconventional ground state, which might be the SSL. Several models have been proposed in the past to explain this behaviour, such as  
the 'order-by-disorder' \cite{Bergman07}, an 'order-by-quenched-disorder' \cite{Savary11} or a 'kinetically inhibited order' \cite{MacDougall11} models.\\
In this work we investigate the relevance of an 'order-by-quenched-disorder' model to our system. It is focused on the well-known experimental fact that materials with the spinel structure could have a significant degree of an A/B antisite inversion. That is, B-atoms occupy the A-site and A-atoms - the B-site. In CoAl$_2$O$_4$ samples studied so far\cite{Tristan05,Tristan08, Zaharko10, Zaharko11, MacDougall11, Hanashima13} the Co/Al inversion $\delta$ varies from 2 to 17\%. The theory\cite{Savary11} predicts that impurities would break the degeneracy of the spiral spin liquid and would induce an ordered magnetic ground state, which is different from the one favoured by the 'order-by-disorder' mechanism. Finally, the 'kinetically inhibited order' model infers that in CoAl$_2$O$_4$ freezing of magnetic domain walls prevents a true long-range order, $\it i.e.$ the thermal energy is too small to overcome the surface energy of domain walls. Thus the size of domains stays 'arrested' at lowest temperatures even in the absence of antisite disorder.\\
For better understanding of the nonconventional magnetic order in the CoAl$_2$O$_4$ system we focus on neutron single crystal diffraction which probes spin-spin correlations and thus should be able to distinguish the models introduced above.  We calculate diffraction patterns for magnetic moment arrangements obtained by the classical Monte-Carlo (MC) simulations varying the $J_2$/$J_1$ ratio, magnetic field and disorder on the A- and B- sites  and compare those with experimental  patterns. Measurements of susceptibility, magnetization and specific heat complement our neutron diffraction results. We conclude that both, the frustration due to the $J_2$/$J_1$ ratio and the presence of the antisite disorder, play a significant role in the observed atypical magnetic ordering. We anticipate that disorder is not random, presumably it happens at $<$111$>$-twin boundaries restricting the propagation of the collinear AF state.\\

\section{MC modelling}{\label{Sec2}}
The Monte-Carlo calculation was performed minimizing the energy of the classical Heisenberg Hamiltonian for spins ${\bf{S}}$: 
\begin{equation}
H = J_{1} \mathop{\sum}_{<ij>} {\bf{S}}_{i} \cdot {\bf{S}}_{j}+ J_{2} \mathop{\sum}_{<<ij>>} {\bf{S}}_{i} \cdot {\bf{S}}_{j}+
D\mathop{\sum}_{i}( {\bf{S}}_{i} \cdot{\bf{u}})^2+g \mu_B B{\sum}_{i}{\bf{S}}_{i} \cdot{\bf{h}}.
\label{eq1}
\end{equation}
\begin{figure}[tbh]
\includegraphics[width=0.33\columnwidth,keepaspectratio=true]{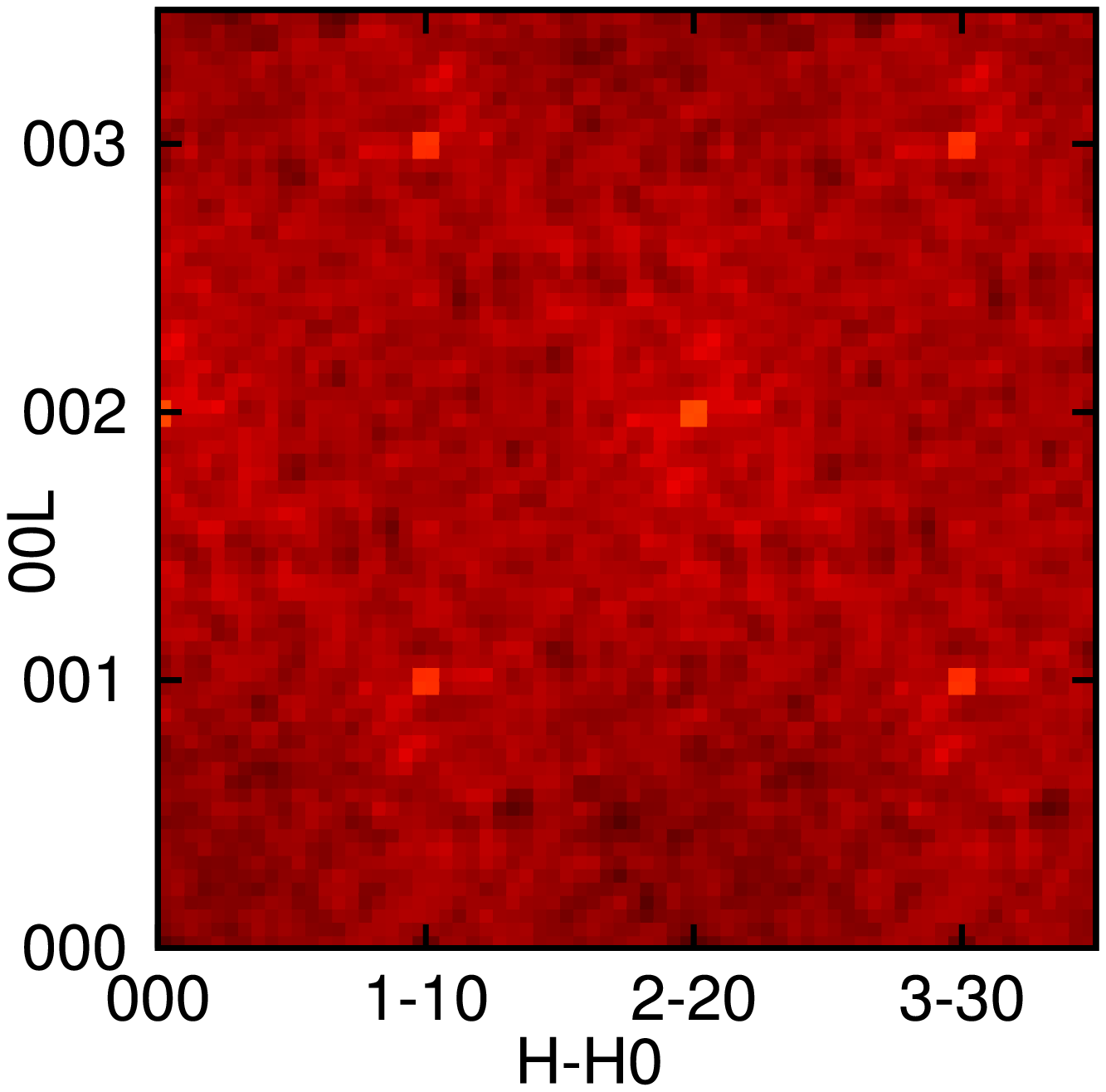}
\includegraphics[width=0.215\columnwidth,keepaspectratio=true]{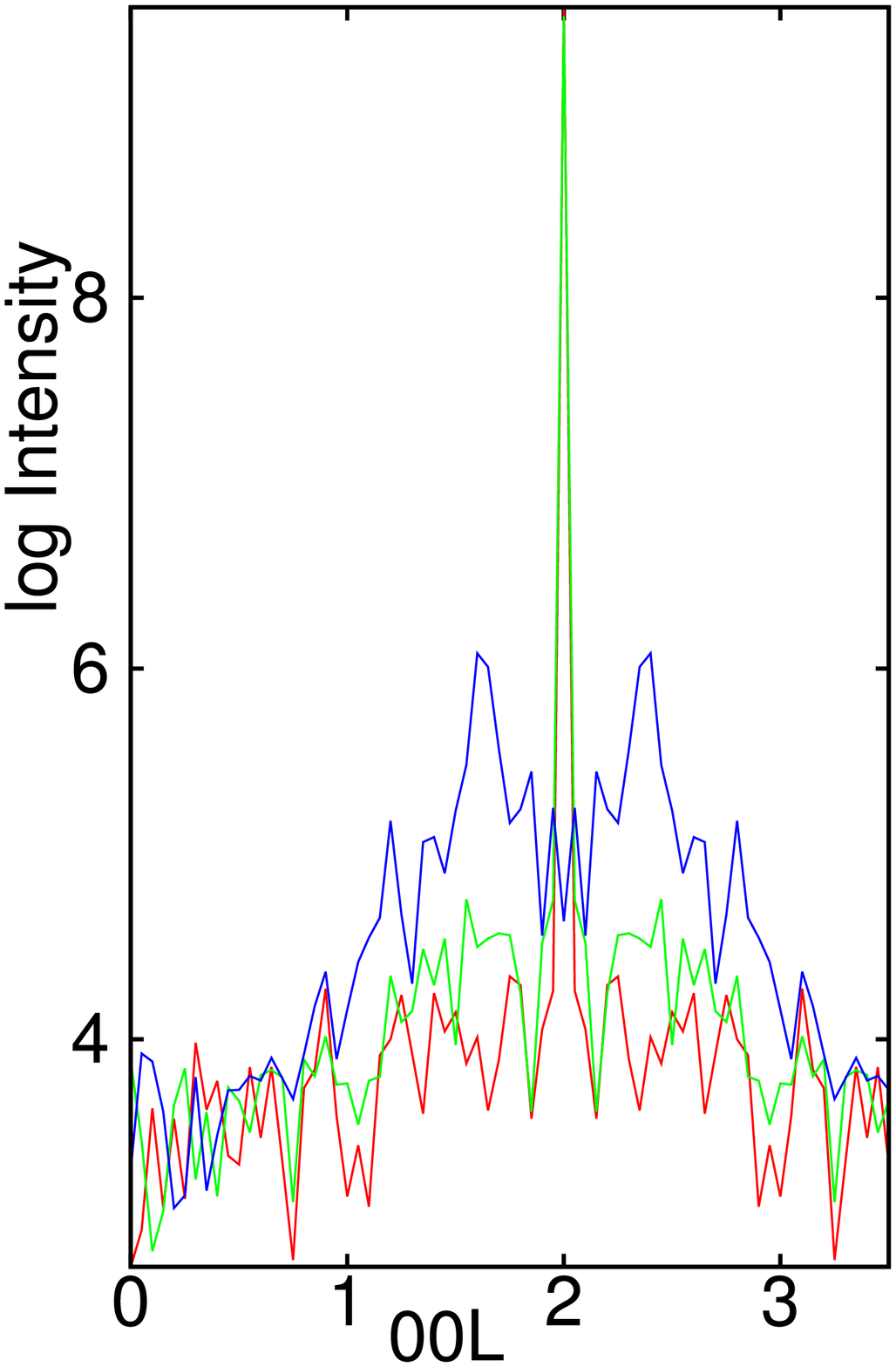}
\includegraphics[width=0.37\columnwidth,keepaspectratio=true]{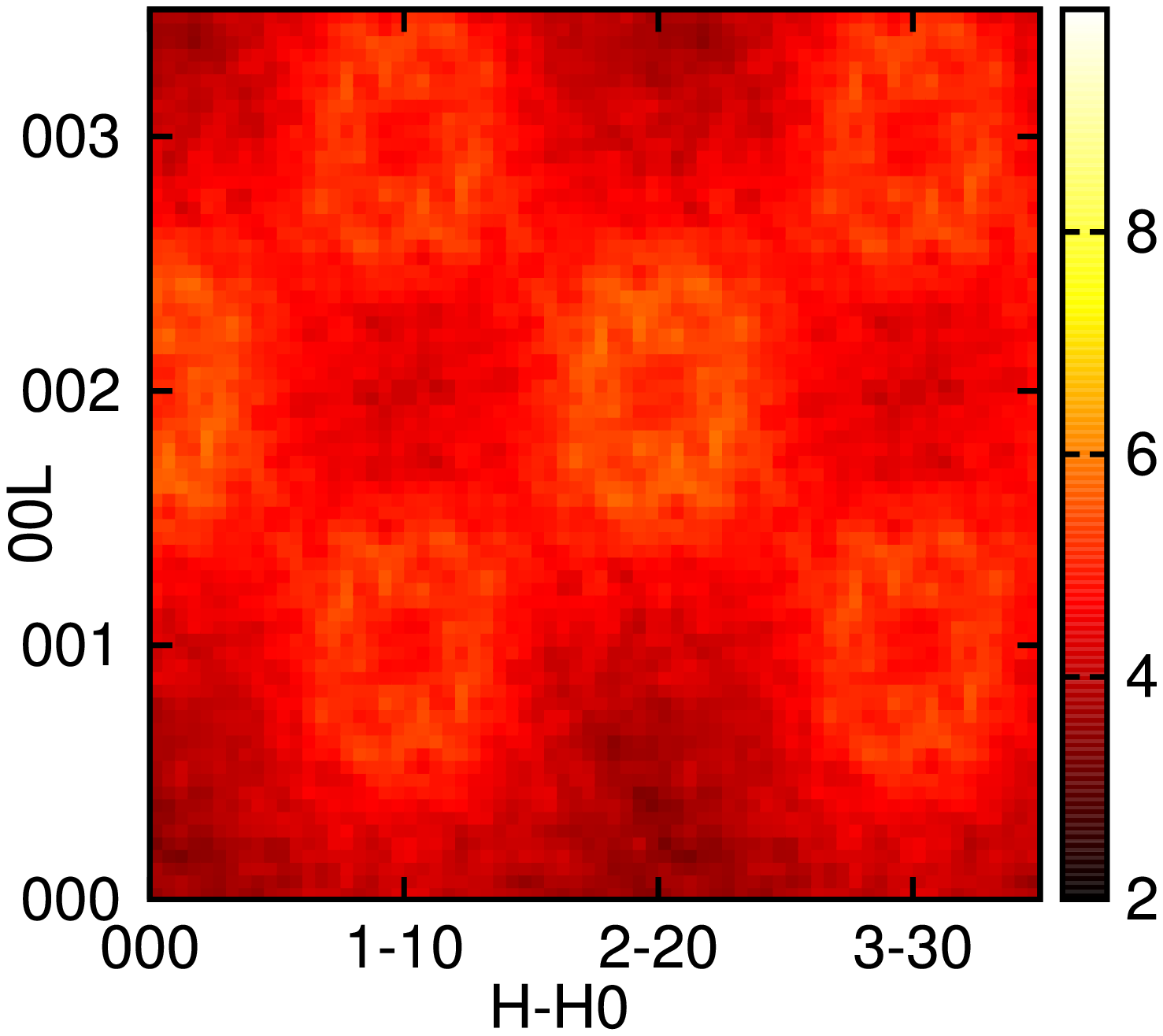}
\caption {Elastic neutron-scattering H-HL maps calculated for the ideal model with $J_2$/$J_1$=0.10 (left), 0.17 (right). Middle: 
The 00L cuts of the H-HL maps with the $J_2$/$J_1$ ratio 0.10 (red), 0.15 (green) and 0.17 (blue). Note the logarithmic scale for the intensity.}
\label{fig1}
\end{figure}
Here $J_{1}, J_{2} >0$ are first- and second-neighbour antiferromagnetic couplings,  $D$ and ${\bf{u}}$ are the magnitude and direction of the single-ion anisotropy term,  $B$ and ${\bf{h}}$ are the magnitude and direction of a magnetic field. 
A cluster of 64000 Co atoms (20x20x20 unit cells) with periodic boundary conditions was considered. This size is a compromise between two conflicting requirements - the cluster
should be sufficiently large to ensure enough resolution in a calculated diffraction pattern, but small enough to spend finite time on a MC calculation. 
The moments were kept constant in magnitude and equal to 1 $\mu_B$
; their direction was changed at random, one at the time. To obtain the ground state 
only energy-decreasing moves were accepted in the final stage. During a MC cycle 10000 moves of each atom were performed to ensure equilibrium conditions. The results were identical to those obtained by temperature annealing. The Fourier transform of the magnetic moment configuration was then calculated for the regions of interest of the reciprocal space. The Q-step was chosen to be 0.05 r.l.u. No binning in the direction perpendicular to the cuts was performed.\\ 
As the main purpose of the modelling was to understand the magnetic behaviour in the CoAl$_2$O$_4$ system, the exchange couplings and disorder parameter close to ones known for our crystal \cite{Zaharko11} were used as the basis, namely $J_1$=9.2 K, $J_2$/$J_1$=0.10, D/J$_1$=0.009 and $\delta$=8\% \cite{note1}. 
Additionally, to disentangle the influence of the parameters, we varied  in reasonable ranges the $J_2$/$J_1$ ratio, amount of random defects and magnetic field strength.
We tried two different single-ion anisotropy directions - $[$111$]$ and $[$100$]$, however, as the anisotropy in this system is small this did not influence the results.
\begin{figure}[tbh]
\includegraphics[width=0.325\columnwidth,keepaspectratio=true]{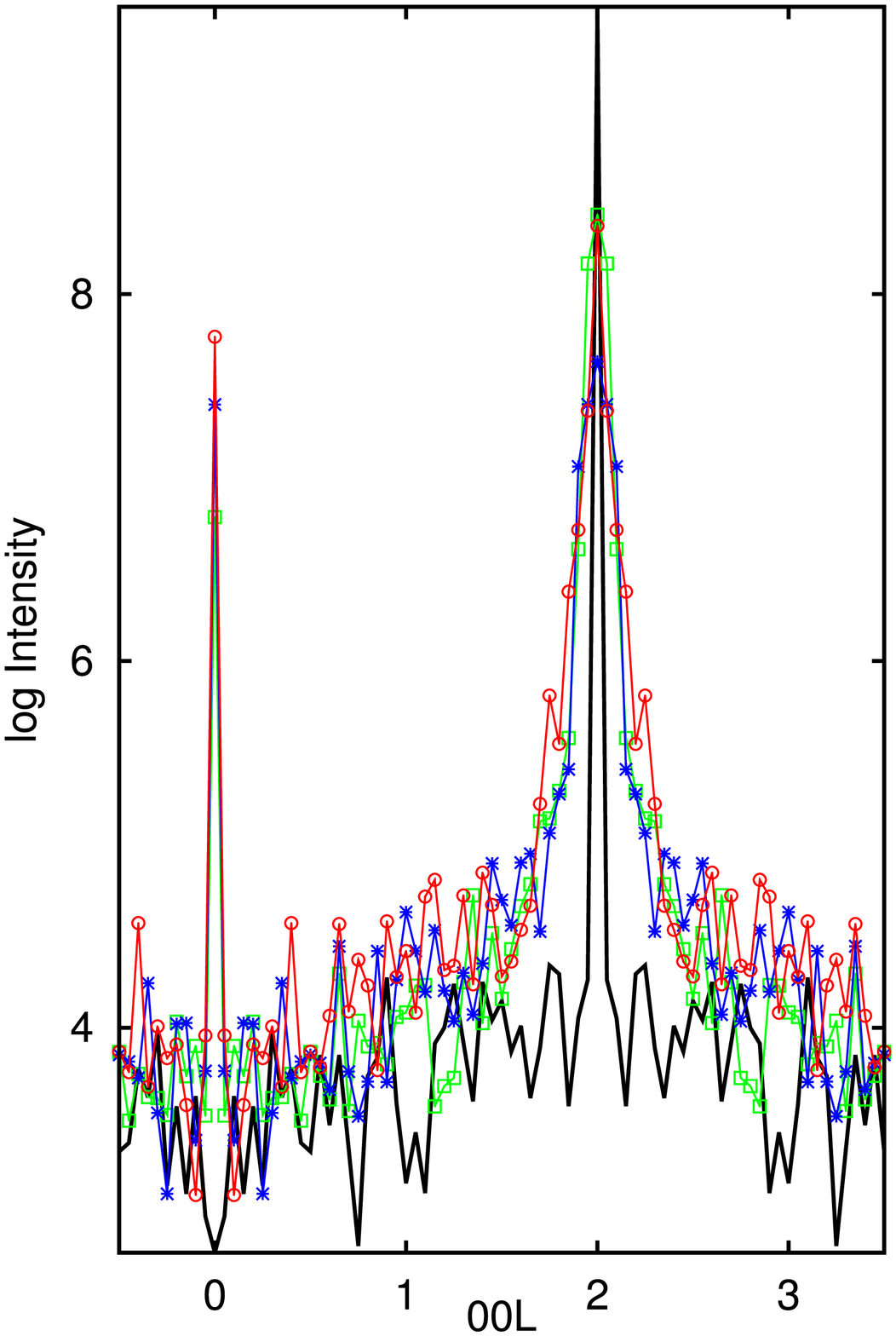}
\includegraphics[width=0.30\columnwidth,keepaspectratio=true]{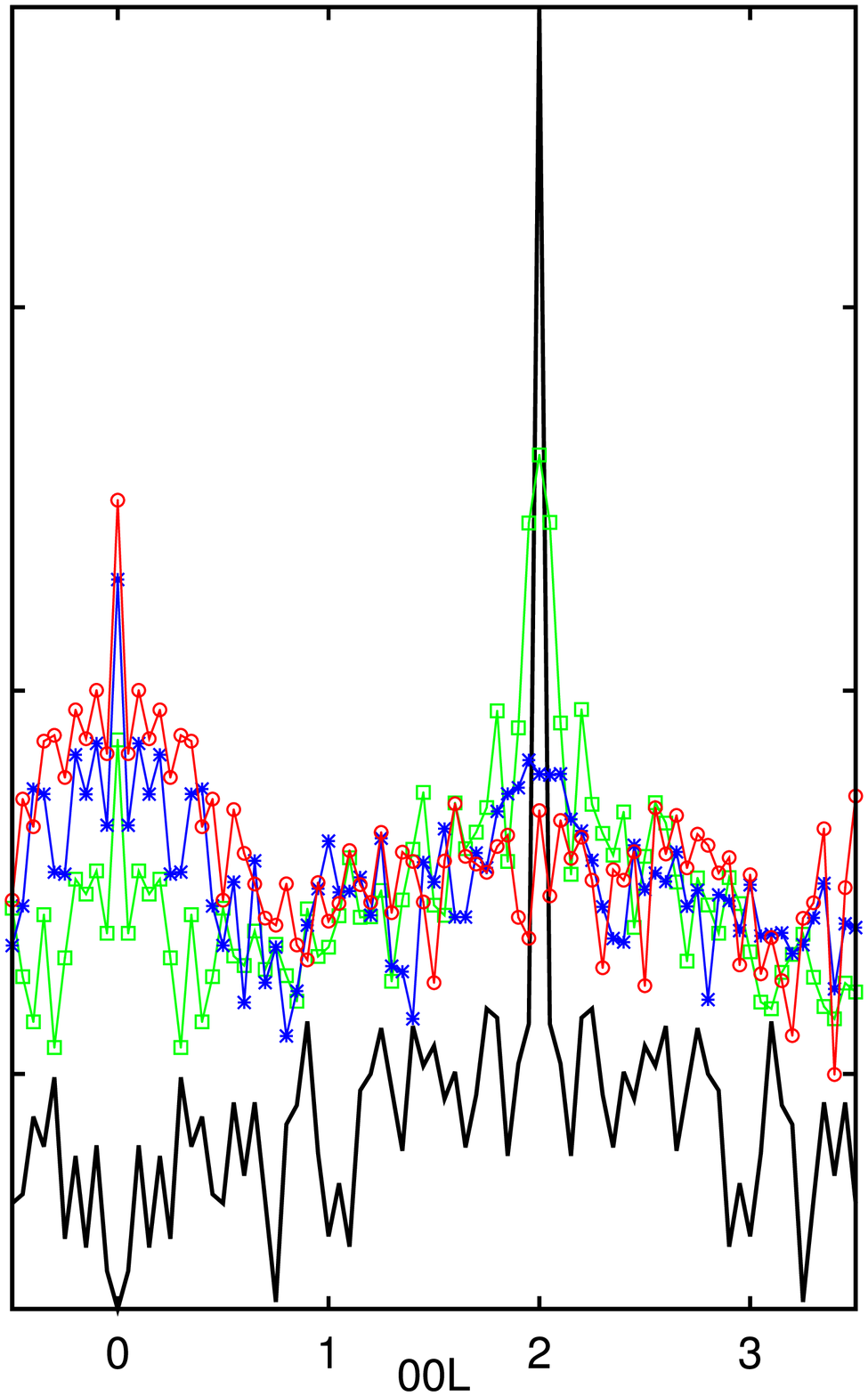}
\caption {00L cuts as a function of A- (left) and B- (right) defect concentration for the model with $J_2$/$J_1$=0.10: 0\% (black), 4\% (green), 8\% (blue), 12\% (red). Note the logarithmic scale for the intensity.}
\label{fig2}
\end{figure}
\\
Representative H$-$HL maps for the ideal (without defects) model with $J_2$/$J_1$ ratios of 0.10 and 0.17 are presented in Fig.~\ref{fig1}. Two main features are apparent: there are sharp magnetic Bragg peaks at the AF (k=0) positions (i.e. 200, 111) and broad diffuse features at the basis of magnetic Bragg peaks extending over several reciprocal lattice units (r.l.u.). The broad diffuse features need the log-scale amplification in single crystal calculated patterns, however summing up this sparse signal at the same sin$\theta/\lambda$ results in a strong diffuse scattering in powder patterns\cite{Zaharko10}.
As can be seen from the 00L cuts in Fig.~\ref{fig1} (middle), the broad diffuse features grow with the increase of $J_2$/$J_1$ (a significant increase happens for $J_2$/$J_1 >$ 0.15), while the Bragg intensity decreases and even a hollow forms at the positions of the magnetic Bragg peaks for $J_2$/$J_1$= 0.17. This reproduces well the 'spiral surfaces' of the free energy calculated by Bergman{\it ~et al.} [\citenum{Bergman07}]. 
\\
\begin{figure}[tbh]
\includegraphics[width=0.27\columnwidth,keepaspectratio=true]{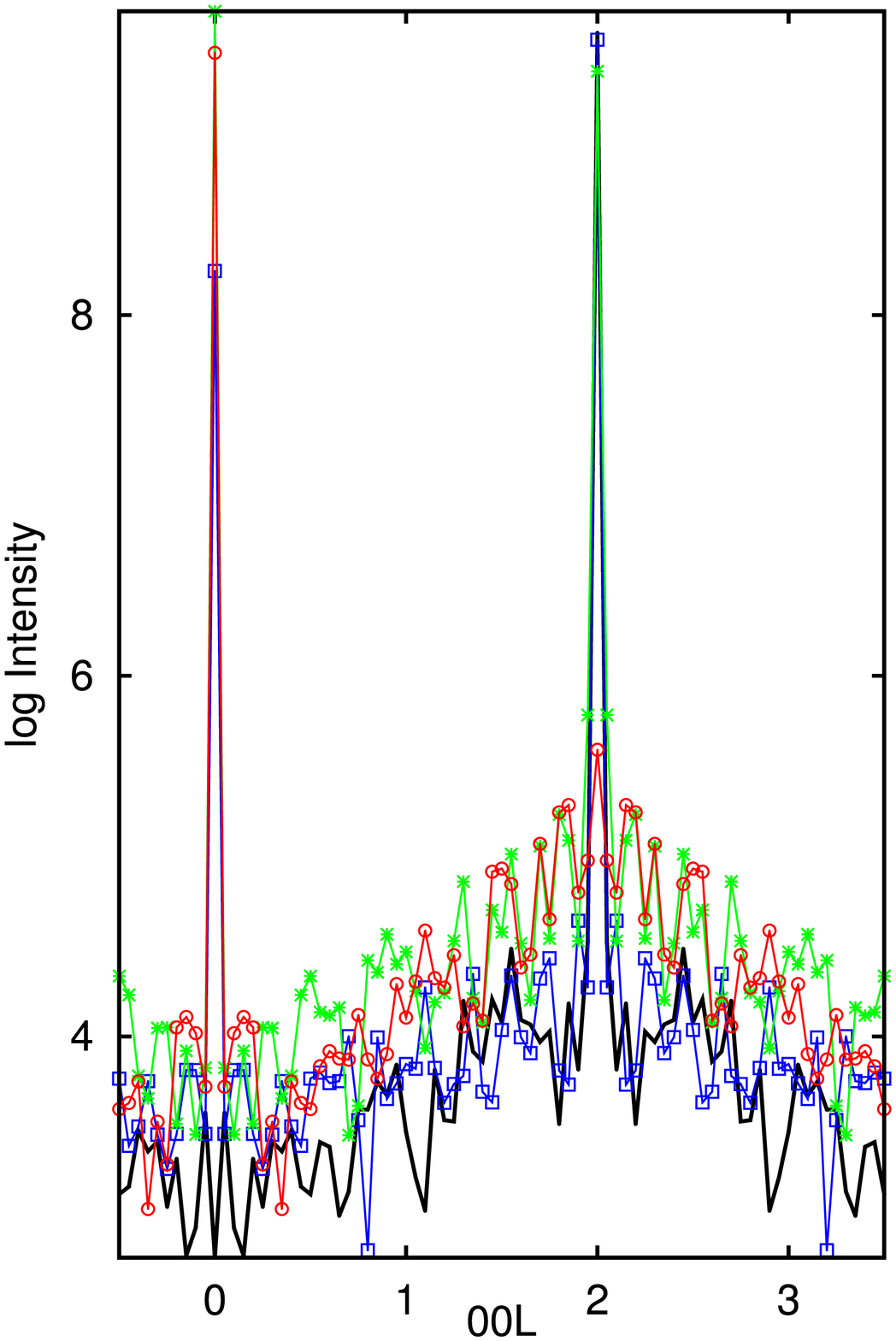}
\includegraphics[width=0.25\columnwidth,keepaspectratio=true]{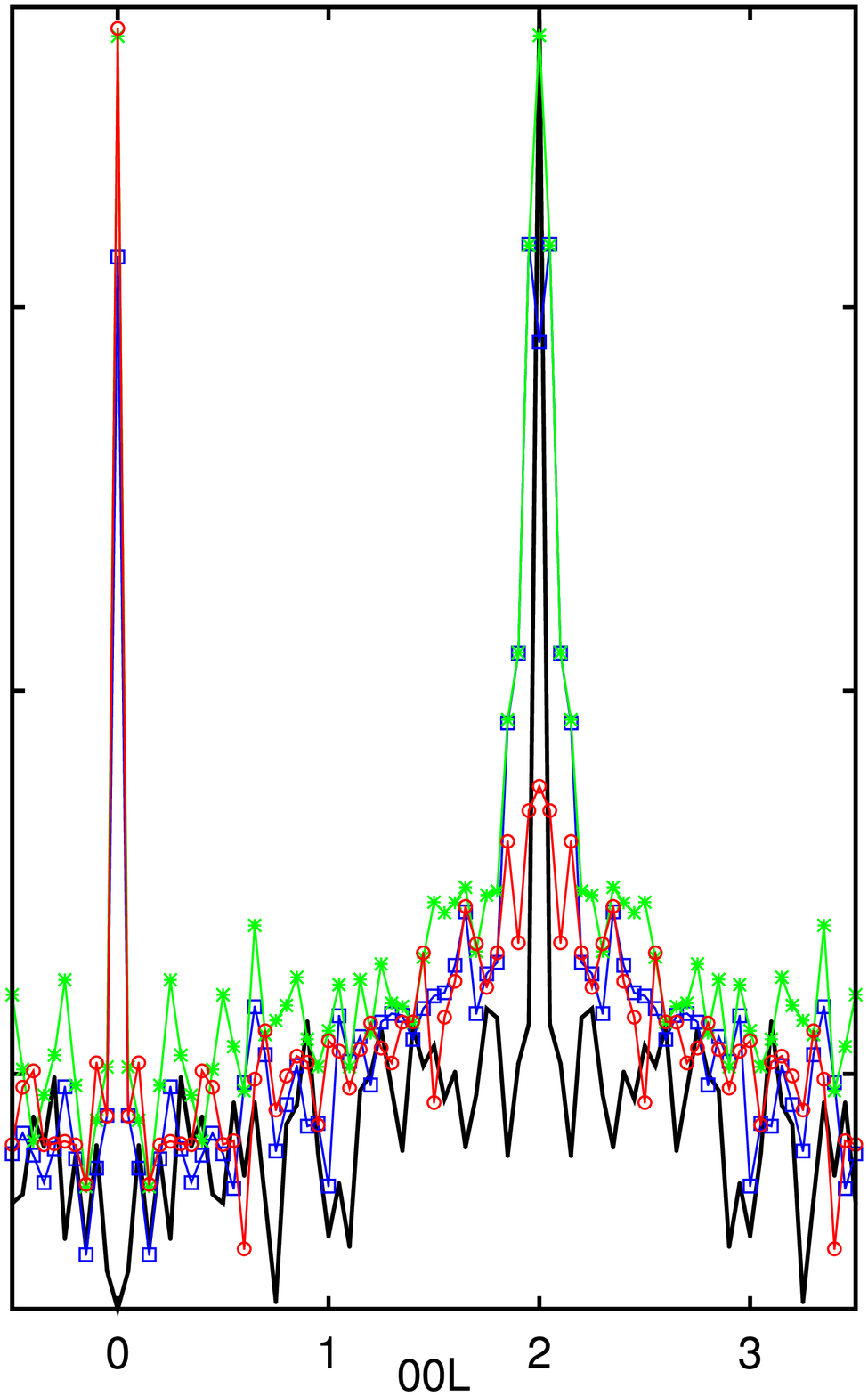}
\includegraphics[width=0.25\columnwidth,keepaspectratio=true]{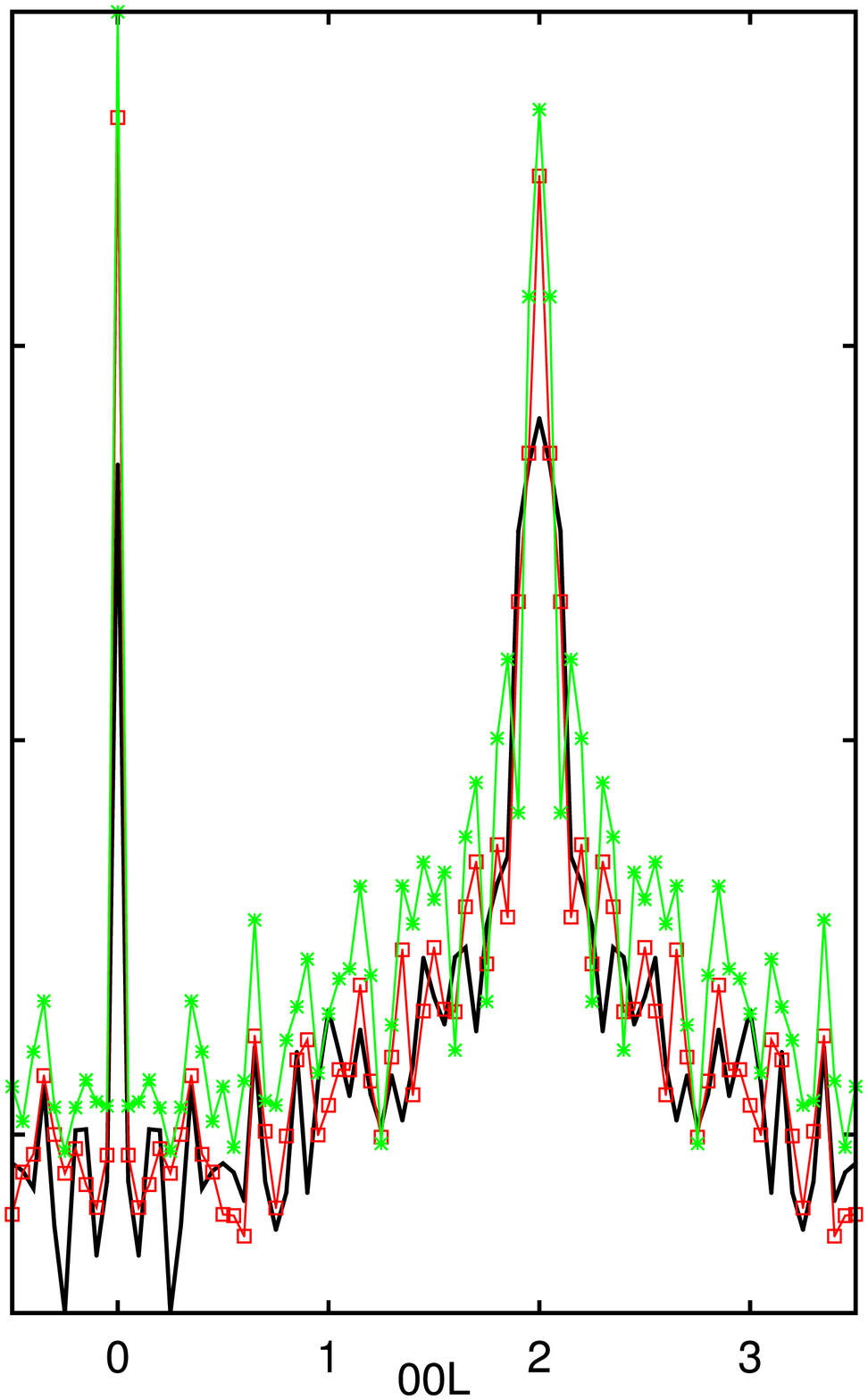}
\caption {00L cuts for zero-field (black), 1.5 T (blue), 4 T (green) and 5 T (red) calculated for the ideal $J_2$/$J_1$= 0.10 model with magnetic fields applied along $[$110$]$ (left) and $[$100$]$ (middle) and for the model with 8\% defects for  4 T with field along $[$100$]$ (red) and $[$110$]$ (green) (right).}
\label{fig3}
\end{figure}
To understand the effect of the Co/Al antisite disorder we considered random defects on the two sites A (Co)- and B (Al) separately. We considered models with random A-defects and B-defects.
First, up to 12\% of magnetic atoms were removed from the A-sites. This significantly changed the intensity distribution, as is presented in Fig.~\ref{fig2} left for the ratio $J_2$/$J_1$=0.10. There appeared sharp ferromagnetic peaks ({\it i.e.} 000, 220), the AF Bragg peaks became wider and the broad diffuse scattering increased.
These tendencies grow with the number of the A-defects. Inspection of atoms in the vicinity of the defects shows that the antiferromagnetic coupling between them is broken and their magnetic moments are aligned along the single-ion anisotropy axis ${\bf{u}}$. This results in the emergence of long-range ferromagnetic correlations at the cost of the antiferromagnetic ones.
\\
Next we inserted up to 12\% of magnetic atoms on the B-sites (see Fig.~\ref{fig2} right). This model, introduced first by Savary{\it et~al.} [\citenum{Savary11}], argues that the coupling between the magnetic B-atom and its six nearest-neighbor A-atoms should be very strong. We adopted this model to the CoAl$_2$O$_4$ case considering the ratio $J_2$/$J_1$=0.10 and we varied the concentration of impurities in the 2 - 12\% range. The results imply that B-defects suppress the AF long-range order much stronger than the A-defects. For concentrations above 4\% antiferromagnetic Bragg peaks vanish. Also at these concentrations  ferromagnetic short-range correlations emerge; they were not detected either for the ideal or A-defect models. 
\begin{figure}[tbh]
\includegraphics[width=0.35\columnwidth,keepaspectratio=true]{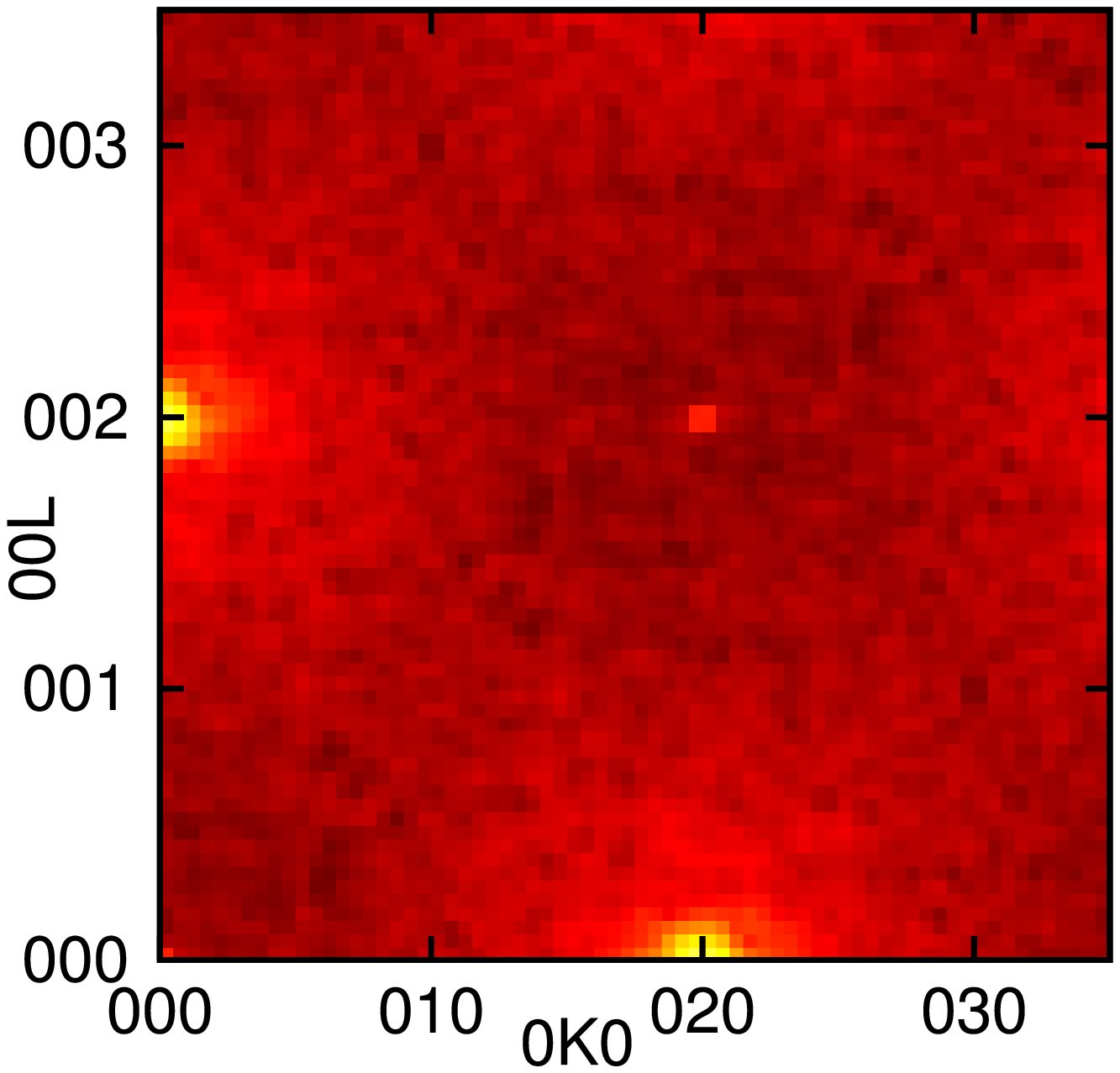}
\includegraphics[width=0.395\columnwidth,keepaspectratio=true]{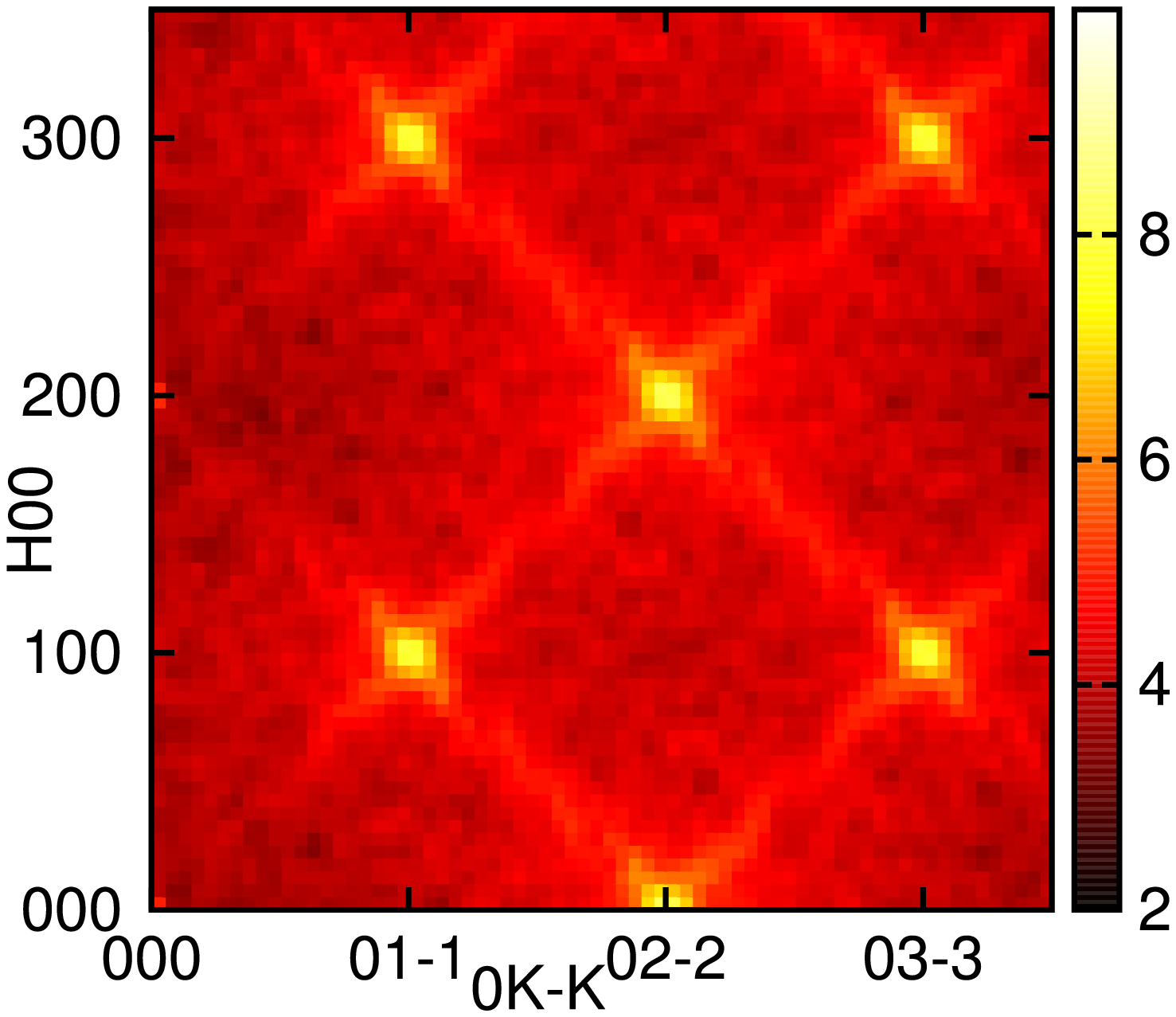}
\caption {Elastic neutron-scattering 0KL (left) and HK-K (right) maps calculated for the ideal model with $J_2$/$J_1$=0.10 with a magnetic field applied along $[$100$]$.}
\label{fig4}
\end{figure}
\\
Further, we wanted to know how a perturbation, such as a magnetic field, would affect diffraction patterns of the ideal and defect models. For that we calculated diffraction patterns for the $J_2$/$J_1$=0.10 ratio without defects and with A-defects. Application of magnetic fields along the principal directions lead to significant changes in magnetic correlations for the ideal model (Fig.~\ref{fig3} left and middle). The common features are: increase of ferromagnetic Bragg peaks, decrease of antiferromagnetic peaks (they vanish at the magnetic field of 5 Tesla) and enhancement of broad diffuse AF features. Still, there are differences.
When magnetic field is applied along $[$110$]$ (Fig.~\ref{fig3} left), diffuse scattering is concentrated around the Bragg peaks, the peaks themselves stay sharp.
For H$\parallel [$100$]$ (Fig.~\ref{fig3} middle) the AF peaks become wider. 
In addition to diffuse scattering located near the AF Bragg peaks (as presented in Fig.~\ref{fig4} left for the 0KL plane with H$\parallel [$100$]$), short-range correlations form diffuse streaks along $<$111$>$.
They are more pronounced in the $<$HK-K$>$ plane for H$\parallel [$100$]$ (Fig.~\ref{fig4} right) compared to the $<$H-HL$>$ plane for H$\parallel [$110$]$ (not shown). This suggests that magnetic field along $[$100$]$ lowers the energy of low-lying states more efficiently than magnetic field along $[$110$]$. 
In clear contrast, when 8\% A-defects are present (Fig.~\ref{fig3} right), the magnetic field enhances all magnetic correlations - not only the ferromagnetic long-range Bragg  contributions and AF diffuse scattering, but also the AF Bragg contributions grow.
\\
Our MC calculations show that magnetic field is a sensitive parameter to identify the relevance of the $J_2$/$J_1$ model and to distinguish the presence or absence of random A- and B-defects. Following these conjectures we decided to perform a neutron single crystal diffraction experiment in the applied magnetic field.
\section{Experimental results}{\label{Sec3}}
\subsection{Bulk measurements}
Prior to neutron diffraction we characterized our single crystal by bulk measurements to compare our results with other numerous studies of the CoAl$_2$O$_4$\cite{Tristan05, Tristan08, Suzuki07, MacDougall11, Hanashima13, Roy13}.
The growth of the CoAl$_2$O$_4$ crystal by the floating zone method was reported by Maljuk {\it et~al.} \cite{Maljuk09}; parts of this crystal were used in the previous neutron scattering experiments \cite{Zaharko11}.
Susceptibility, magnetization and specific heat measurements presented here were performed on
a part of this crystal with the mass of 12.5 mg using a Quantum Design Squid Vibrating Sample Magnetometer (SVSM) and a Quantum Design PPMS, respectively.\\
\begin{figure}[tbh]
\includegraphics[width=0.30\columnwidth,keepaspectratio=true]{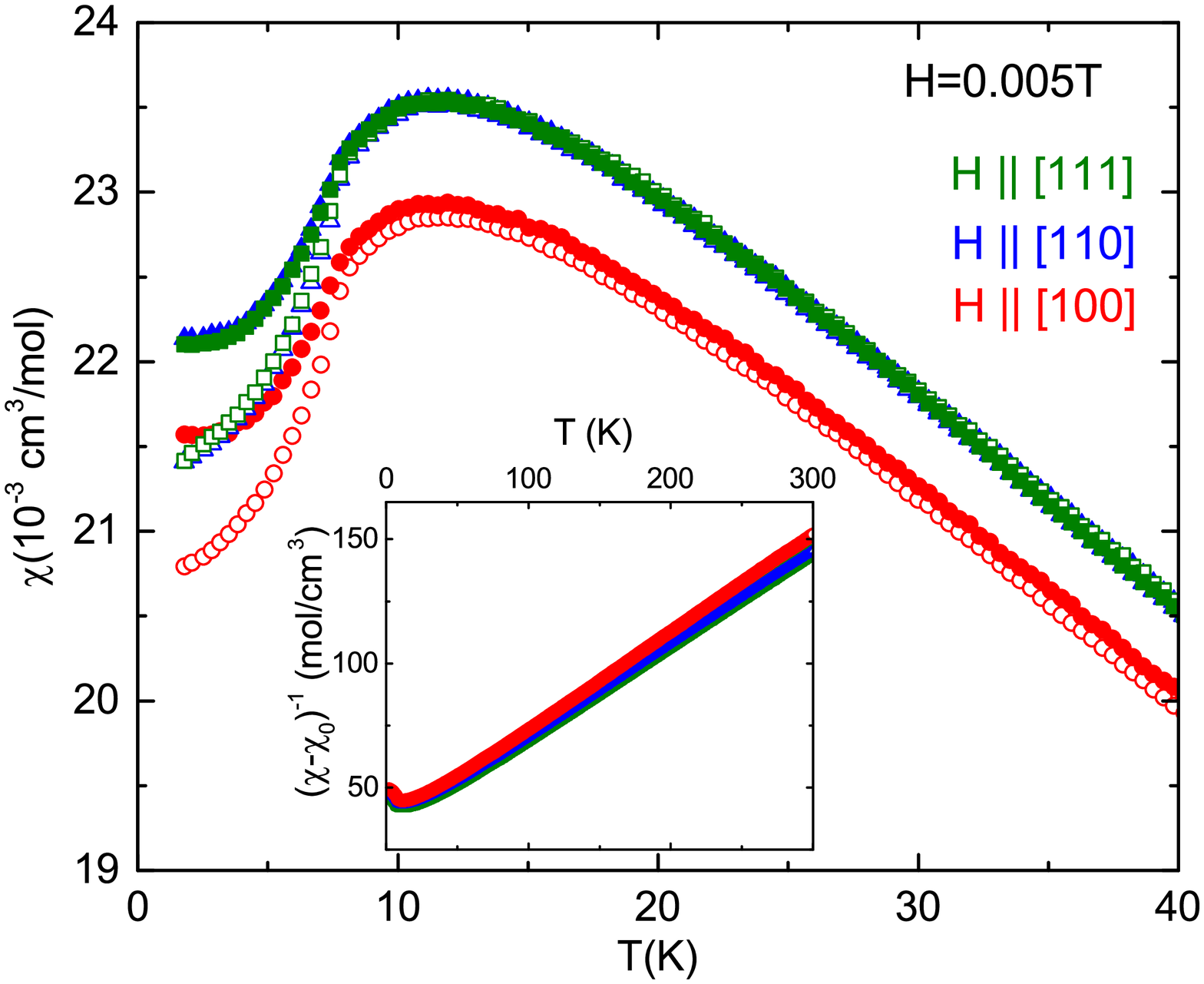}
\includegraphics[width=0.30\columnwidth,keepaspectratio=true]{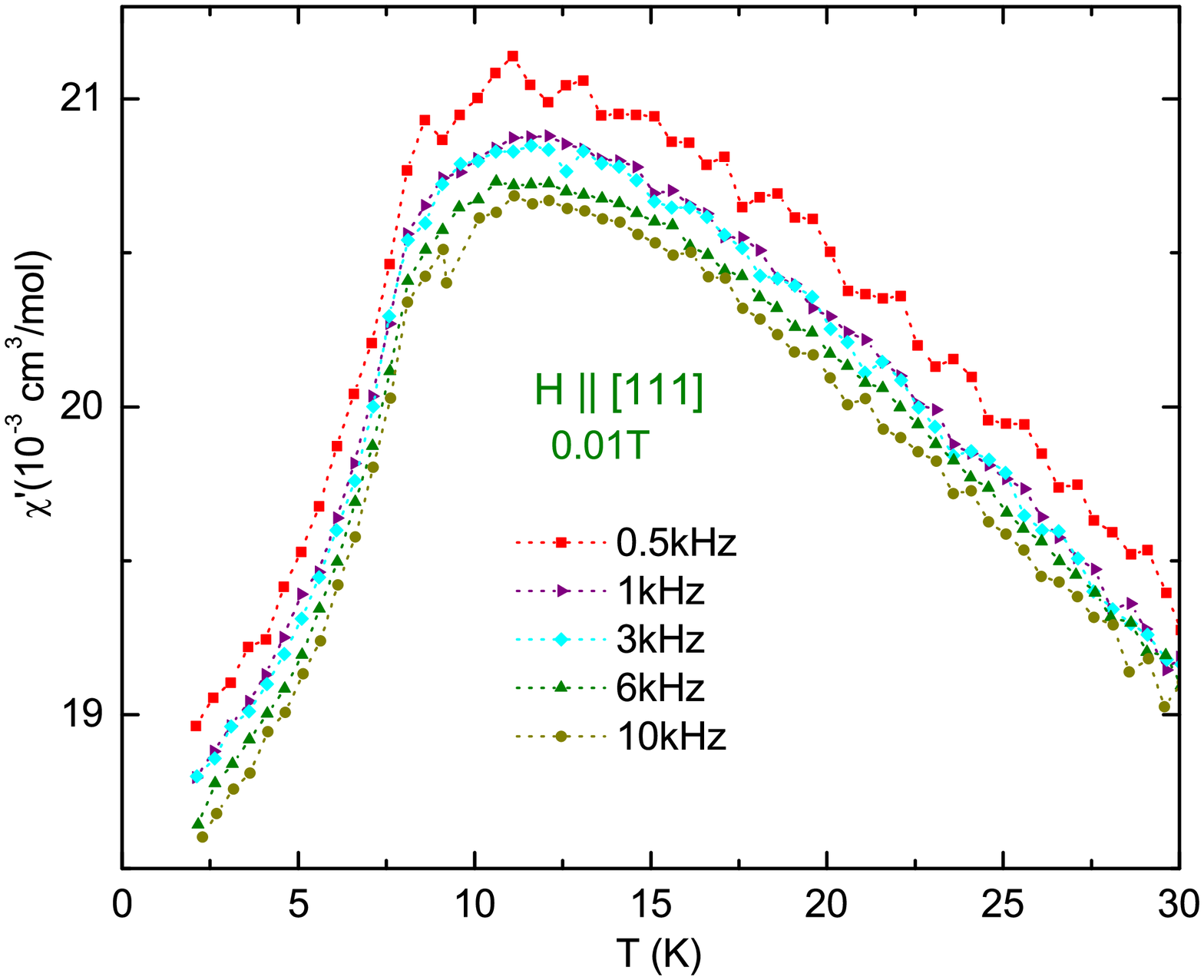}
\includegraphics[width=0.30\columnwidth,keepaspectratio=true]{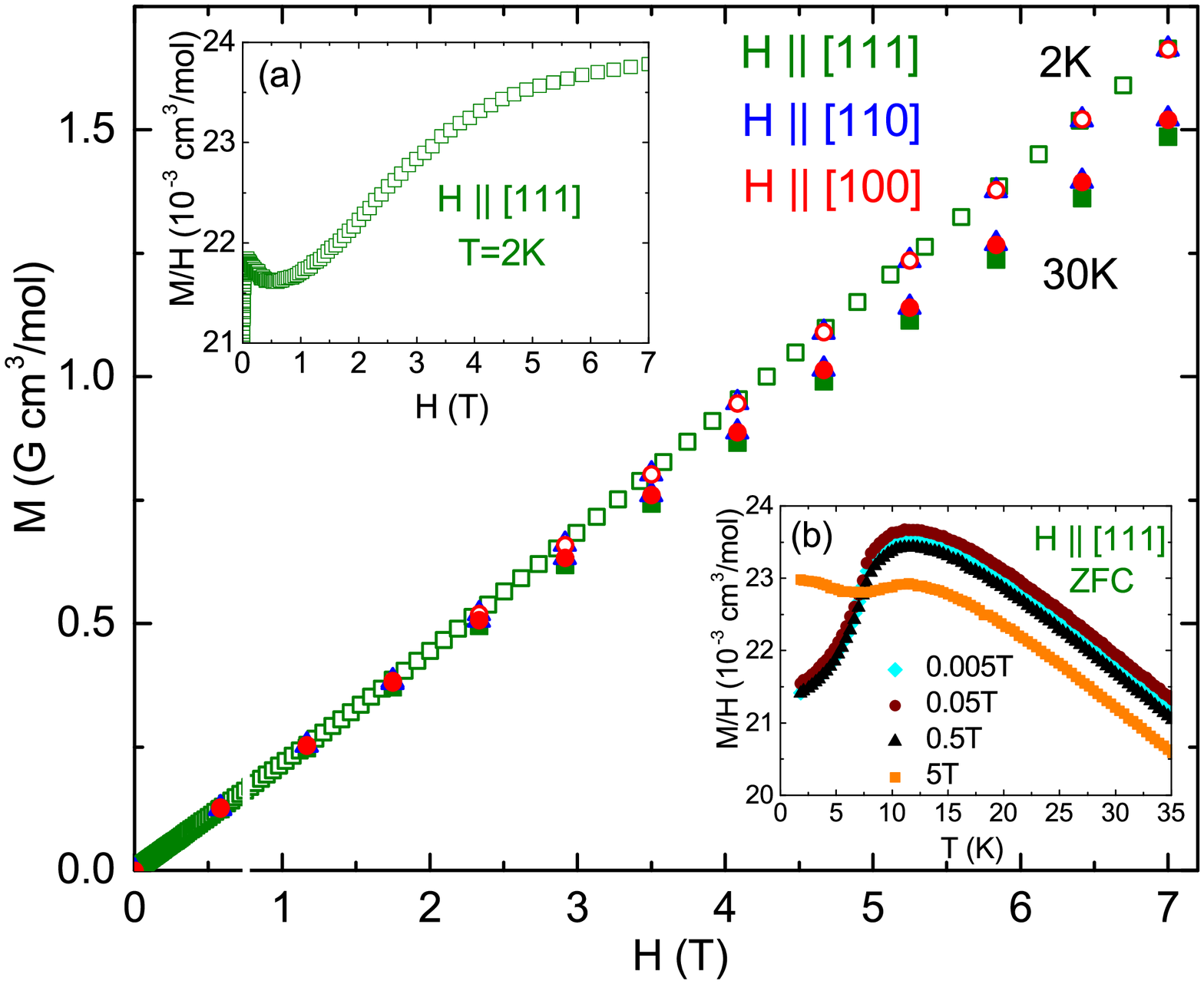}
\caption {Left: Temperature dependence of magnetic susceptibility $M/H$ measured with a DC field $H$=0.005 T. Solid symbols represent field cooled and open symbols - zero field cooled data. Green - magnetic field applied along the $[$111$]$ direction, blue - $[$110$]$, red - $[$100$]$. Inset: inverse susceptibility for three principal directions. Middle: Temperature variation of AC magnetic susceptibility $\chi'$ for different frequencies with a DC field $H$= 0.1 T along $[$111$]$. Right: Magnetization at 2 K (open symbols) and 30 K (solid symbols). Insets: a) Change of $M/H$ with H$\parallel[$111$]$. b) Variation of $M/H$ with temperature at different strength of magnetic field along $[$111$]$.}
\label{fig5}
\end{figure}
The low temperature zero field cooled (ZFC) and field cooled (FC) susceptibility measured with a DC magnetic field $H$= 0.005 T applied along three principal directions is shown in Fig. \ref{fig5} left. The high-temperature part, above 100 K, follows a Curie-Weiss law with a Weiss temperature $\theta_{CW}$= -85.2 K. The determined effective magnetic moment 4.60 $\mu_B$/f.u. and g-factor 2.37 are in good agreement with other studies \cite{Roy13}.
A broad hump at $\sim$11 K indicates building up of short range antiferromagnetic correlations. Below T$^*\sim$8 K a small splitting between the ZFC and FC susceptibilities  without a pronounced cusp is observed. Remarkably, there is no frequency dependence of this anomaly in the AC susceptibility (Fig. \ref{fig5} middle). These observations are in excellent agreement with the systematic study of antisite disorder in the Co$_{1-\delta}$Al$_\delta$(Al$_{1-\delta}$Co$_{\delta}$)$_2$O$_4$ system by Hanashima  {\it et~al.}\cite{Hanashima13}, which attributes such behaviour to a spin liquid state (samples of Suzuki {\it et~al.}\cite{Suzuki07} and Roy {\it et~al.}\cite{Roy13} belong to this regime as well). Hanashima  {\it et~al.}\cite{Hanashima13} report a crossover between spin liquid (SL) and spin glass (SG) regimes with a threshold at the 8\% inversion level. Samples belonging to the SG regime have a cusp in the DC susceptibility at the ZFC and FC splitting point and the real part of the AC susceptibility is frequency dependent (samples of Tristan {\it et~al.} \cite{Tristan05,Tristan08, Hanashima13} belong to this range). Presumably in the SL regime defects only perturb the degenerate ground state, while in the SG regime they start to dominate magnetic behaviour.
As both, the ZFC and FC splitting in the susceptibility and the broad magnetic peaks in neutron diffraction in our crystal \cite{Zaharko11}, appear
at T$^* \sim$8 K, we identify this anomaly as the emergence of an unconventional ordered magnetic state.\\
Isothermal magnetization curves for our crystal with the field along three principal  axes for 2 K and 30 K are presented in Fig. \ref{fig5} right. At 30 K magnetization is linear
with the field as expected for a paramagnet, while at 2 K the behavior is nonlinear (as emphasized in the inset (a) for $[$111$]$ direction). Such nonlinearity would be expected for reorientation of magnetic domains, however
$M/H$ is identical regardless of the field direction thus suggesting that magnetic anisotropy in this system is very weak.\\
\begin{figure}[tbh]
\includegraphics[width=0.38\columnwidth,keepaspectratio=true]{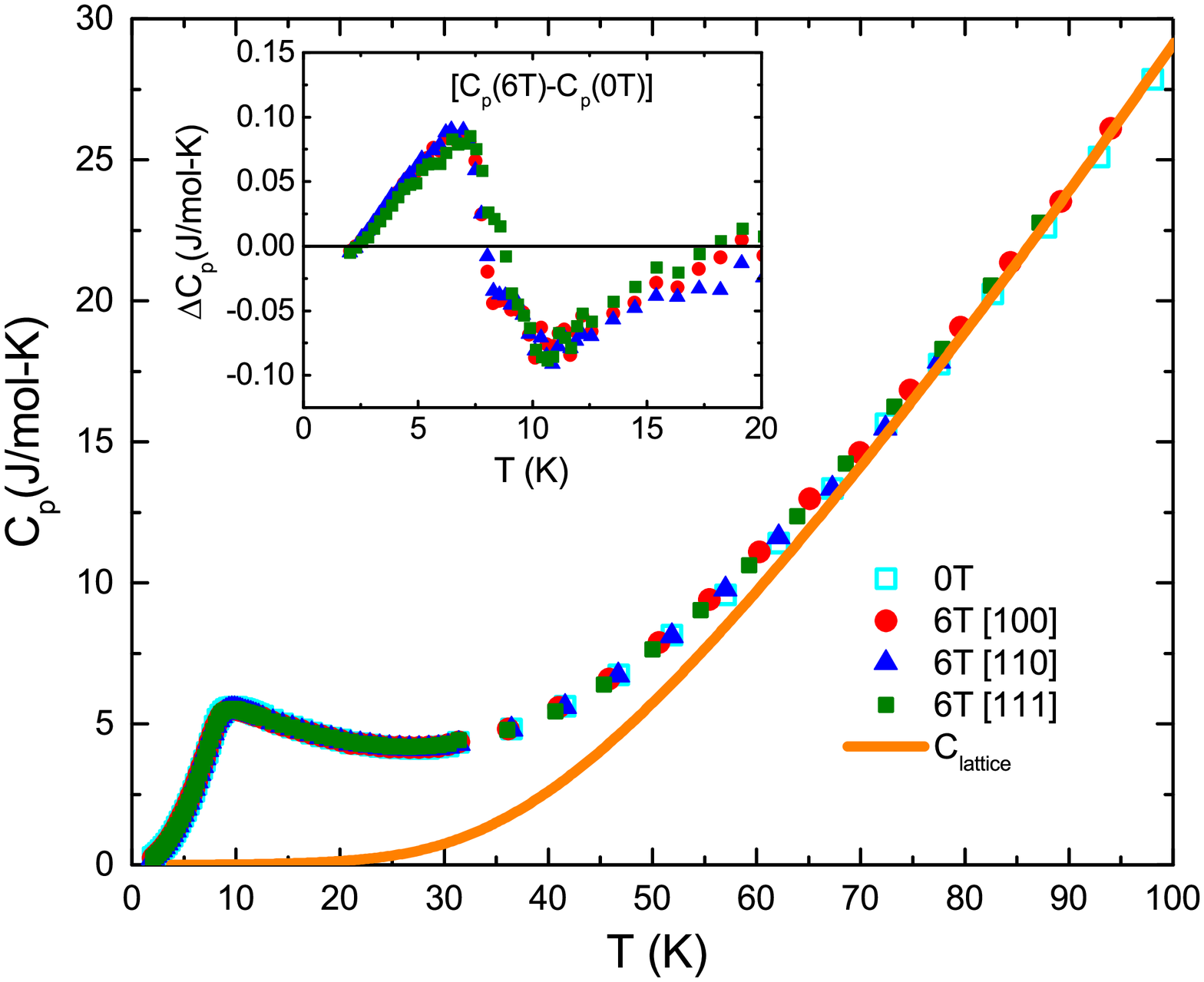}
\includegraphics[width=0.38\columnwidth,keepaspectratio=true]{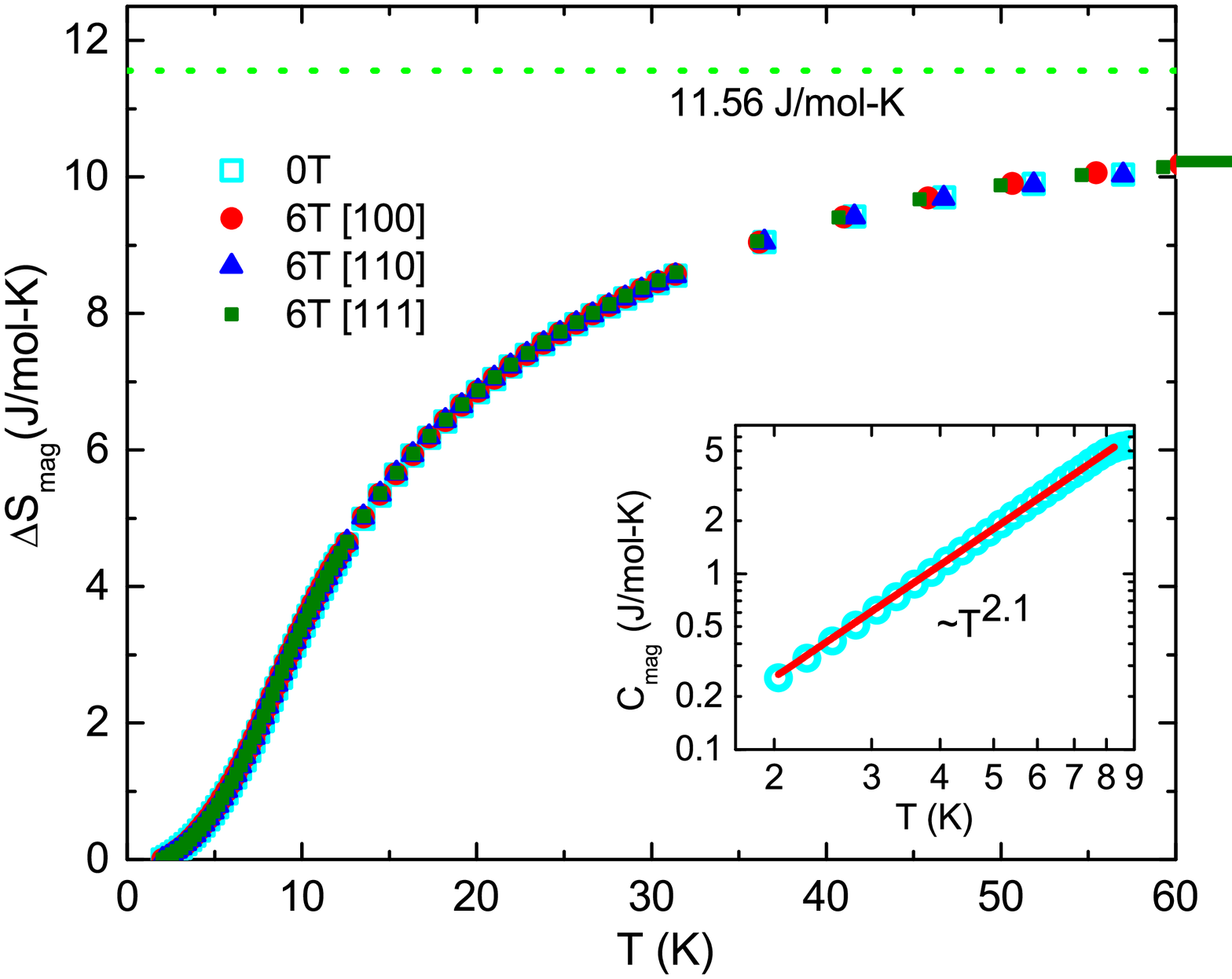}
\caption {Left: Temperature variation of heat capacity C$_{P}$ measured in zero and 6 T magnetic field applied along three principal directions: red $[$100$]$, blue $[$110$]$, green $[$111$]$. Heat capacity of ZnAl$_{2}$O$_{4}$ (non-magnetic analog) taken from Ref.[\citenum{Tristan05}] is also plotted. Inset: Difference of C$_{P}$ measured in 6 T and without field. Right: Magnetic entropy $\triangle\mathrm{S}_{mag}$ calculated from these data. The dashed horizontal line
indicates the high temperature limit of $\triangle\mathrm{S}_{mag}$= 11.56 J/mol K for S=3/2. Inset: Magnetic
contribution to the heat capacity C$_{mag}$ fitted to a power law with ${\alpha}$=2.1.}
\label{fig6}
\end{figure}
The temperature dependence of the heat capacity measured in zero field and in the applied magnetic field of 6 T along the three
principal directions are very much alike (Fig. \ref{fig6} left). For all measurements the crystal was cooled in zero field and then a magnetic field was applied. A broad peak located at $\sim$9.8 K in zero field slightly shifts to lower temperature when magnetic field is applied (inset of Fig. \ref{fig6} left). The change though tiny and independent of the field direction, suggests that only few degrees of freedom are involved and their response to the magnetic field averages throughout the sample. To obtain the magnetic contribution $C_{mag}$ we subtracted the heat capacity of the 
crystal lattice modelled by the Debye approximation \cite{Debye}.
As shown in the inset of Fig. \ref{fig6} right, $C_{mag}$ follows the power law T$^{\alpha}$ with ${\alpha}$=2.1 (red solid line) within the temperature range $2-8$ K. In other samples reported in literature the power laws vary from ${\alpha}$=2.5, ${\alpha}$=2.23 for the SL regime samples \cite{Suzuki07, Roy13}
to ${\alpha}$=2 for the GS sample \cite{Tristan08}. These ${\alpha}$-values should be compared to the power law expected for a classical antiferromagnetic order ($C_{mag} \sim$ T$^{3}$),  for the spiral surface ($J_2$/$J_1$) model ($C_{mag} \sim$ T$^{7/3}$)\cite{Bergman07} and a linear behavior typical for spin glass systems ($C_{mag} \sim$ T). The agreement is in
favor of the $J_2$/$J_1$ model and the presence of defects seems to affect the entropy very little. Notably, the saturation of the magnetic entropy $\triangle\mathrm{S}_{mag}$ evaluated by integrating $C_{mag}/T$ over T (Fig. \ref{fig6} right) is significantly lower than the expected high-temperature limit for $S=3/2$ systems
(11.56 J/molK). Presumably a significant part of entropy remains in the system up to higher temperatures.

\subsection{Neutron diffraction}
\begin{figure}
\includegraphics[width=0.30\linewidth, trim=0 0 0 0, clip=true]{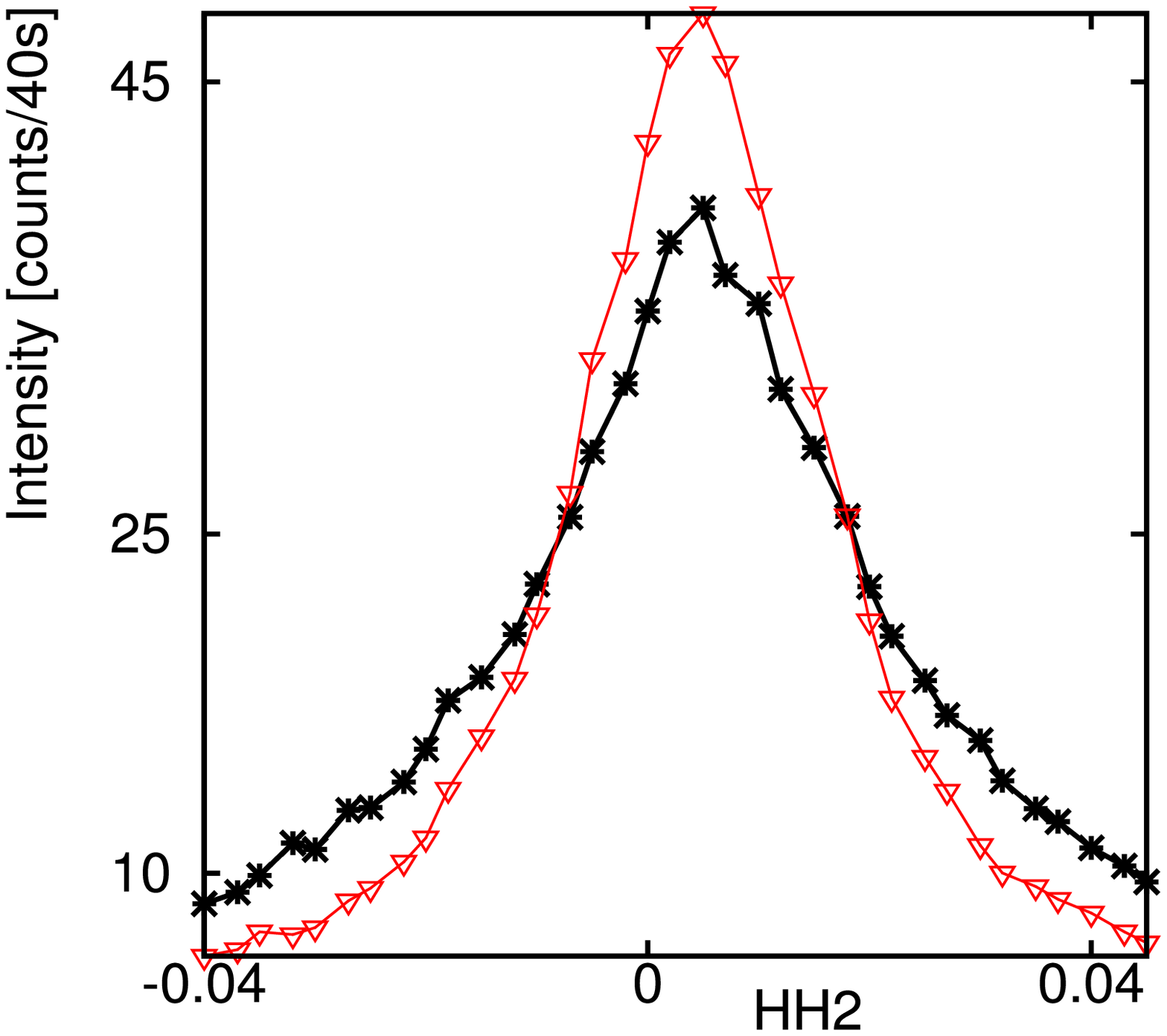}
\includegraphics[width=0.27\linewidth, trim=0 0 0 0, clip=true]{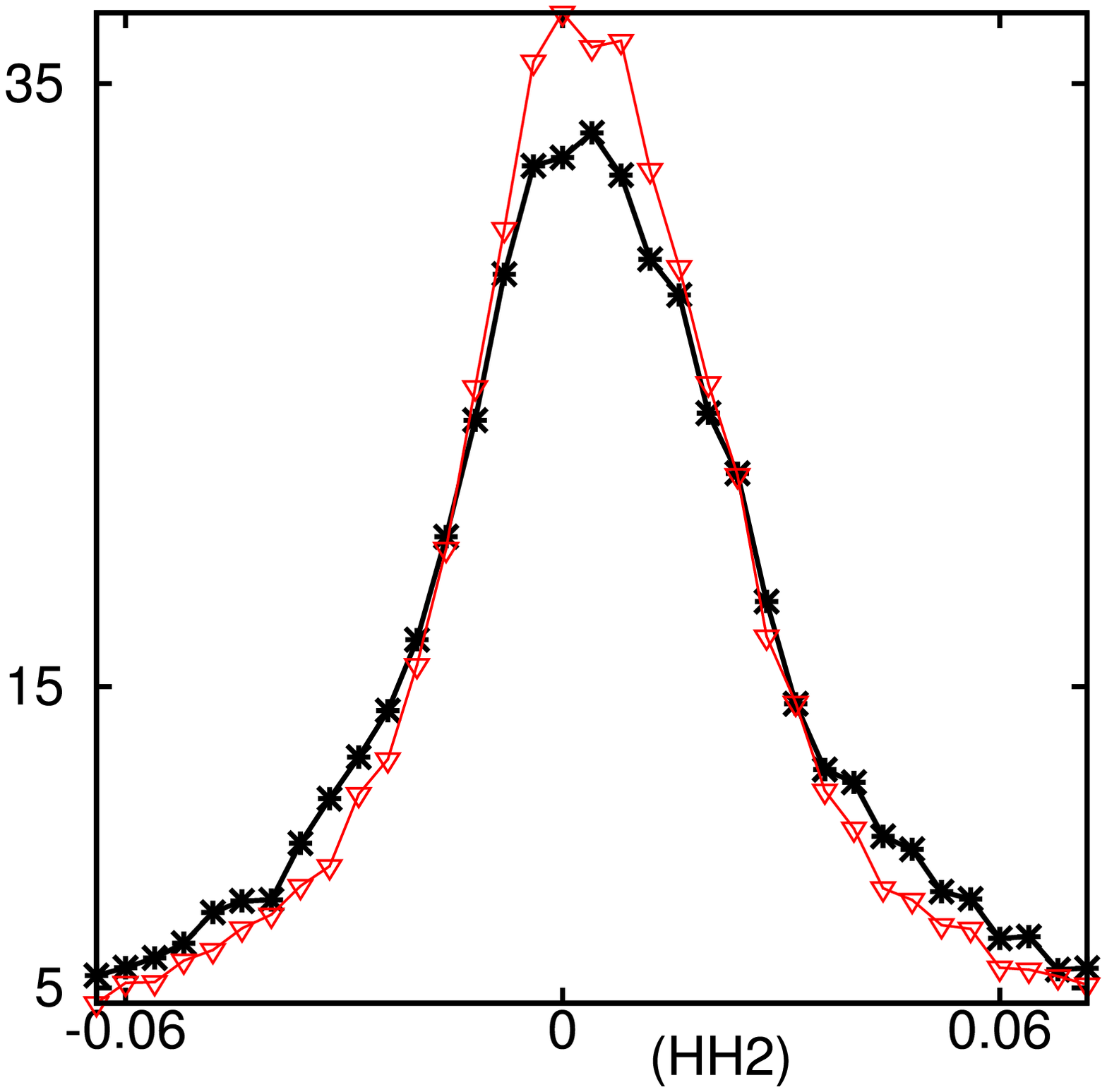}
\includegraphics[width=0.27\linewidth, trim=0 0 0 0, clip=true]{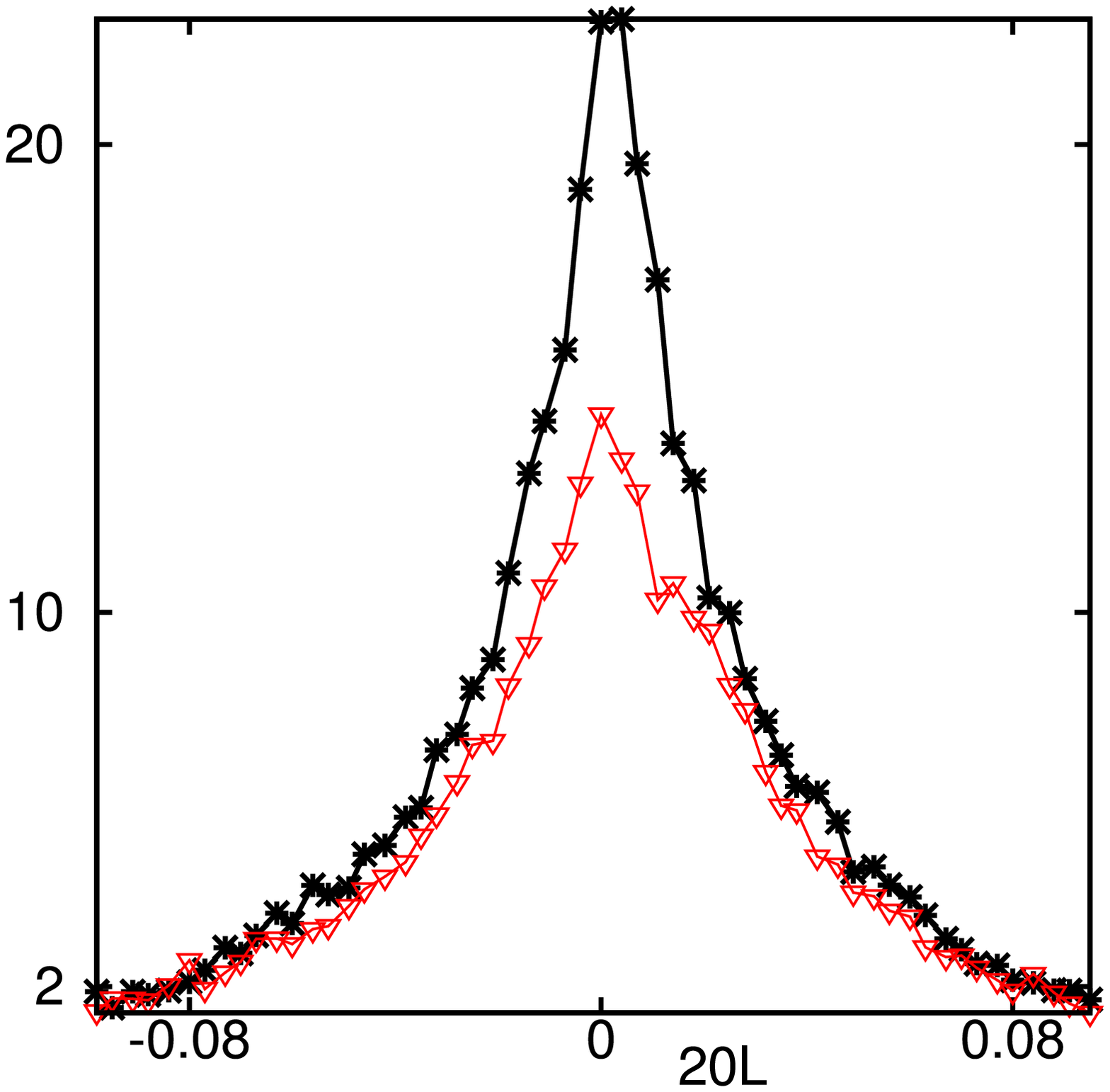}
\caption {$<$200$>$ reflections of CoAl$_{2}$O$_{4}$ crystal measured at 1.8K without (0 T, black) and with (5 T, red) magnetic field applied along $[$110$]$ (left), $[$111$]$ (middle) and $[$010$]$ (right).}
\label{fig7}
\end{figure}

Neutron single crystal experiments were performed on a 0.2755 g crystal cut from the CoAl$_2$O$_4$ ingot at the base temperature of 1.8 K on the single crystal diffractometer TriCS at SINQ, Villigen.
Magnetic field was applied vertically along the principal directions $[$010$]$,  $[$110$]$ and  $[$111$]$ by reorienting the crystal in a MA06 cryomagnet. 
The neutron wavelength of 1.18 \AA~and a 80' collimator in front of a $^3$He detector were used.
In zero-field, the magnetic peaks of the $<$00$2>$ family are significantly ({\it c.a.} 3 times) broaden (Fig.~\ref{fig7}) compared to the instrumental resolution.
This resembles broad peaks of the MC model with A-defects and not the features of the ideal MC model
(sharp peaks and broad diffuse scattering extending over several r.l.u.).
However, contrary to the predictions of the models with random defects, no magnetic intensity is present at the ferromagnetic (nuclear) positions in zero field.
\\
In a magnetic field, the intensity and shape of magnetic peaks change and these changes strongly depend on the field direction. The magnetic field applied along the $[$110$]$ direction (Fig.~\ref{fig7} left) reduces the width of the (002) magnetic reflection and significantly (15\%) increases the peak intensity at the maximum. However, the total intensity remains almost the same. 
This observation suggests that long-range correlations grow at the cost of short-range correlations perpendicular to (002) being in accord with the increase of the AF long-range correlations in the MC model with A-defects.\\
For H$\parallel[$010$]$ (Fig.~\ref{fig7} right) the intensity on the (200) magnetic peak maximum decreases by 37\% and the total intensity decreases by 21\%. 
This contradicts the MC model with A-defects, which predicts an enhancement of the AF long-range correlations.
The field applied along the $[$111$]$ direction (Fig.~\ref{fig7} middle) changes the peak maximum and the total intensity of the magnetic peaks (200), (00-2) and (020) very little. This also contradicts the MC model with A-defects.\\ 
It should be noted that our experimental findings cannot be explained within the conventional model of a multidomain AF structure either. That model postulates the presence of magnetic domains, each of them having the same magnetic structure with a fixed magnetic moment direction along one of complementary preferred easy axes (for example, octet of the $<$111$>$ axes). When a magnetic field is applied, domains with magnetic moments orthogonal to the field direction grow at the cost of domains with unfavourable moment axes. In our case, however, it seems that magnetic moments rotate orthogonal to the applied magnetic field but not necessarily into domains of the complementary easy axis. Such soft magnetic behaviour can explain the increase of the (002) intensity maximum for H$\parallel [$110$]$ and its reduction for the H$\parallel [$010$]$ magnetic field direction.\\

\section{Discussion}
\begin{figure}
\includegraphics[width=0.70\linewidth, trim=0 0 0 0, clip=true]{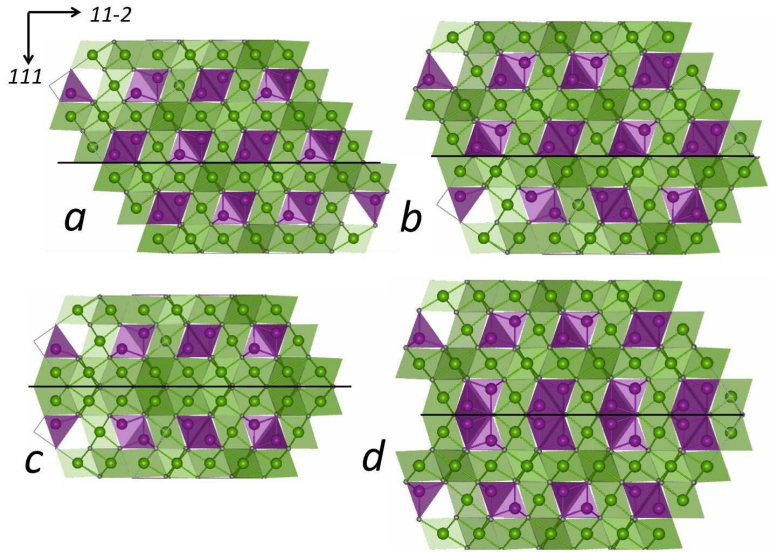}
\caption {The $[$1-10$]$ projection of the AB$_2$O$_4$ spinel structure with a) ideal arrangement of Co$^{+2}$ A-tetrahedra (violete) and Al$^{+3}$ B-octahedra (green) and $<$111$>$-twin boundaries preserving the cation composition (b)
or resulting in the B-rich (c), A-rich (d) regions.}
\label{fig8}
\end{figure}

{\label{Sec4}}
In Sec. \ref{Sec2} we calculated the single-crystal neutron diffraction patterns based on the ideal 'order-by-disorder' model and extended the
simulation to the 'order-by-quenched-disorder' models with random A- and B- defects. In Sec. \ref{Sec3} our experimental data on our CoAl$_2$O$_4$
single crystal are presented. Our macroscopic (susceptibility, magnetization and specific heat) and microscopic (single crystal neutron diffraction in zero and in the applied magnetic field) experimental findings confirm that the magnetic order in the frustrated diamond lattice antiferromagnet CoAl$_2$O$_4$ is unconventional. 
Comparison of bulk properties of our crystal with those of other studied samples identifies our crystal as located in the spin liquid regime\cite{Hanashima13} and its magnetic properties are isotropic on the macroscopic scale. This is definitely not the case on the microscopic level as testified by neutron diffraction.
Single crystal patterns reveal that changes of magnetic correlations with the applied field are anisotropic. The patterns are similar to those calculated from the Monte-Carlo models, but none of the considered models explains the data completely. In zero field, the measured peaks are broadened, as the model with A-defects predicts, however, the anticipated ferromagnetic contribution is missing. 
In addition, while the experimentally observed increase of long-range AF correlations for H$\parallel [$110$]$ is in accord with this model, its decrease for H$\parallel [$100$]$ contradicts the predictions. Furthermore, the presence of random B-defects should induce broad ferromagnetic features, but they are not observed in the experiment.\\
Thus, the magnetic behaviour in the studied material is not captured by simplistic models supposing random disorder on the A- and B-sites considered so far.
One of the possible reasons of the failure could be different electronic states of the defect atoms as the assumed ones. For example, the Co$^{+2}$ ions inserted into the octahedral B-sites are assumed to be in the same spin state as in the A-site. However DFT calculations \cite{Tielens06} suggest the low spin configuration for the Co$^{+2}$ ions in the octahedral coordination for the inversion levels up to $\delta$=50\%. Another possible reason of disagreement could be a complex morphology of defects which we assumed to be random. 
Spinels are known to form complicated twin structures \cite{Klapper12}, which lead to stacking faults of the $<$111$>$ planes. Such faults have been observed in MgAl$_2$O$_4$ by transmission electron microscopy (TEM) \cite{Carter87}. 
Starting from the ideal spinel structure shown in Fig.~\ref{fig8} (a) a stacking fault develop at the (111) plane indicated by the black solid line. At such twin boundary the cation composition can be either preserved (Fig.~\ref{fig8} b), or it can become B(Al)-rich (Fig.~\ref{fig8} c) or A(Co)-rich (Fig.~\ref{fig8} d). If the distribution of stacking faults is similar in all  eight directions within the $<$111$>$ family, macroscopic magnetic properties of a crystal would remain also isotropic.\\ 
Clearly, the frustration due to the $J_2$/$J_1$ ratio is the basic ingredient of the magnetic behaviour of CoAl$_2$O$_4$. This is the case even when the $J_2$/$J_1$ ratio is below {\eight}. The $J_2$/$J_1$ model correctly predicts the temperature evolution of specific heat,
the wave vector of magnetic ordering and the presence of low-lying excited states with $\bf{q}$= $<$111$>$. However, not all experimental observations emerge from the $J_2$/$J_1$ physics. Both, the difference in ZFC/FC susceptibility and the broadening of the AF peaks in neutron diffraction result from the antisite disorder. Comparison of neutron patterns  from the MC modelling and the experiment prompts that these models are too simplistic. TEM would be the key method for a systematic study of the microstructure and its influence on magnetic correlations in this material reconciling contradicting isotropic macroscopic and anisotropic microscopic observations.\\
\begin{acknowledgments}
This work was performed at SINQ, Paul Scherrer Institute, Villigen, Switzerland with financial support of the Swiss National Science Foundation (project SCOPES IZ73Z0 152734/1).
\end{acknowledgments}


\begin{thebibliography}{99}
\bibitem{note1} Preparing this manuscript we revisited the x-ray diffraction refinements and found that the degree of inversion in our crystal is actually lower, 5\%.
\end{thebibliography}
\end{document}